\documentclass[draftcls,onecolumn,11pt]{IEEEtran}

\usepackage{amsmath,amsfonts}

\usepackage{cite}
\usepackage{algorithm}
\usepackage{algpseudocode}
\usepackage{enumerate}
\usepackage{color}
\usepackage{setspace}
\usepackage{balance}
\usepackage{url}

\usepackage[pdftex]{graphicx}
\graphicspath{{./}}%
\DeclareGraphicsExtensions{.jpg,.png,.pdf}
\usepackage[font={footnotesize}]{caption}
\usepackage[caption=false,font=footnotesize]{subfig}

\newcommand{\bi}{\begin{itemize}}	
\newcommand{\ei}{\end{itemize}}
\newcommand{\bn}{\begin{enumerate}}	
\newcommand{\en}{\end{enumerate}}
\newcommand{\bc}{\begin{center}}
\newcommand{\ec}{\end{center}}
\newcommand{\be}{\begin{equation}}
\newcommand{\ee}{\end{equation}}
\newcommand{\bea}{\begin{eqnarray}}
\newcommand{\eea}{\end{eqnarray}}
\newcommand{\ben}{\begin{equation*}}
\newcommand{\een}{\end{equation*}}
\newcommand{\beqa}{\begin{eqnarray}}
\newcommand{\eeqa}{\end{eqnarray}}
\newcommand{\btabu}{\begin{tabular}}
\newcommand{\etabu}{\end{tabular}}

\begin{document}
\title{\LARGE \bf
Distributed Algorithm for Collision Avoidance at Road Intersections in the Presence of Communication Failures}

\author{\IEEEauthorblockN{Vladimir Savic, Elad M. Schiller, and Marina Papatriantafilou \vspace{0.5em}\\}
\IEEEauthorblockA{Dept. of Computer Science and Engineering, Chalmers University of Technology, Sweden\\
Emails: \{savicv, elad, ptrianta\}@chalmers.se
}
}
\maketitle

\begin{abstract}
Vehicle-to-vehicle (V2V) communication is a crucial component of the future autonomous driving systems since it enables improved awareness of the surrounding environment, even without extensive processing of sensory information.
However, V2V communication is prone to failures and delays, so a distributed fault-tolerant approach is required for safe and efficient transportation. In this paper, we focus on the intersection crossing (IC) problem with autonomous vehicles that cooperate via V2V communications, and propose a novel distributed IC algorithm that can handle an unknown number of communication failures. Our analysis shows that both safety and liveness requirements are satisfied in all realistic situations. We also found, based on a real data set, that the crossing delay is only slightly increased even in the presence of highly correlated failures.

\end{abstract}

\section{Introduction}\label{sec:intro}\footnote{Copyright (c) 2017 IEEE. Personal use of this material is permitted. However, permission to use this material for any other purposes must be obtained from the IEEE by sending a request to pubs-permissions@ieee.org. The original version of this paper is submitted to IEEE Intelligent Vehicles Symposium (IV'2017).}
Future \textit{autonomous vehicles} will enable safer, more efficient and more comfortable transportation \cite{Wymeersch2015}. They will be equipped with a wide range of sensors \cite{Mukhtar2015}, such as Global Positioning System (GPS) receivers, radars, lidars, cameras, and inertial measurement unit (IMU). In addition, they are expected to have radios for wireless communication \cite{Hafner2013, casimiro2013} that would be used to exchange all relevant information with nearby vehicles and the infrastructure. This would facilitate increased awareness of surrounding environment, including the distant objects out of the sensing horizon. Moreover, the vehicles would be able to optimize their trajectory using the sensory information and the future positions from nearby vehicles. This work focuses on fully autonomous vehicles with a vehicle-to-vehicle (V2V) communication unit and a minimal set of sensors (e.g., GPS and IMU) required for safe and efficient transportation. Our approach is extendable and can include, for redundancy sake, other sources of sensory information, that would increase the robustness to unforeseen failures, and would detect the passive objects that are not able (or not willing) to communicate.

In particular, we focus here on intersections since these parts of the roads account for almost half of all accidents \cite{Azimi2013}. The intersections are typically managed by traffic lights and stops signs, but these systems would cause an excessive delay with autonomous vehicles. On the other hand, since wireless communication is prone to failures and delays, a centralized intersection manager is not a desirable solution. We rather consider a \textit{distributed} method in which the vehicles need to agree, via V2V communication, on the order in which they should cross the intersection. Our solution advances the state of the art (see Section \ref{subsec:relwork}) because we ensure collision avoidance in the presence of an \textit{unknown} number of communication failures, and without a significant increase in the crossing delay. 

The remainder of this paper is organized as follows. In Section \ref{sec:back}, we provide the background and the related work for the problem at hand, and in Section \ref{sec:smodel} we formulate the problem and provide models for intersection crossing (IC), including the position-related information and the message format. Then, in Section \ref{sec:distalg}, we provide our novel algorithm for distributed IC in the presence of communication failures, and, in Section \ref{sec:numres}, we analyse numerically the expected delay caused by these failures. Finally, in Section \ref{sec:conc}, we summarize our results and provide suggestions for future work.

\section{Background}\label{sec:back}

Autonomous driving is a multi-discipline problem \cite{Wymeersch2015}, mainly consisting of sensor fusion, communication and control units that interact between each other. Sensor fusion unit is responsible for acquiring, processing and fusing all available data. This data is obtained from a wide variety of on-board sensors, and also received from the nearby vehicles. The final estimates are then used to feed the control unit, which is responsible to handle the vehicle, i.e., ensure that the vehicle is moving according to the desired velocity and acceleration. Finally, the communication unit allows vehicles to exchange the relevant data which are then used, in combination with local data, to generate an appropriate control action. Since the communication range is typically larger than the sensing range, the vehicles will have more time to make an appropriate decision. However, wireless communication is prone to failures and delays, so a robust solution is required for a reliable communication. We focus on this problem in this paper.

\subsection{Related Work}\label{subsec:relwork}
We overview here the state-of-the-art on IC algorithms for autonomous and semi-autonomous vehicles.

In \cite{Mukhtar2015}, the authors provide a survey on vehicle detection techniques, with a focus on vision-based detection. The sensors are first classified into two groups: active (such as lasers, radars and lidars) and passive (such as cameras, and acoustic sensors), and then compared to each other in terms of range, cost and other features. The radar is considered as the best active sensor, since it provides long-range ($>150 m$) real-time detection even under bad weather (e.g., foggy, rainy) conditions. On the other hand, a radar is not able to estimate the shape of the object, which can be done with lidar, a costly alternative. These problems encouraged authors to focus on passive sensors such as cameras. Cameras are low-cost sensors, able to provide a very precise information about the objects. However, their main drawback is a high complexity of data processing, low range during nights, and sensitivity to weather conditions. Note that authors did not consider any kind of communication between vehicles, that would resolve some of the sensors' problems.

In \cite{Hafner2013}, the authors use V2V for decentralized and cooperative collision avoidance for semi-autonomous vehicles, in which the control is taken from the driver once the car enters a critical area. The algorithm is tested using  vehicles equipped with: differential GPS (DGPS), IMU, dedicated short-range communication (DSRC) unit, and an interface with actuators. Their solution aims to compute appropriate throttle/brake control to avoid entering the capture area, in which no control action can prevent a collision. The estimation of longitudinal displacement, velocity and acceleration is performed using Kalman filtering. This estimation takes into account a bounded communication delay found experimentally. Their experimental results showed that all collisions are averted, and that the algorithm does not introduce a significant delay.

The work in \cite{Azimi2013} develops reliable and efficient intersection protocols using V2V communication. The proposed solutions are able to avoid deadlocks and vehicle collisions at intersections. The protocols are fully distributed since they do not rely on any centralized unit such as intersection manager. The autonomous vehicles are equipped with a similar set of sensors as in \cite{Hafner2013}, and also a DSRC unit for V2V communication. The vehicles interact with each other using standardized basic safety messages (BSM) adapted for intersection crossing. The proposed protocols are tested using AutoSim simulator/emulator, which utilizes a real city topography. The results showed that the proposed protocols outperform the traditional traffic light protocols in terms of trip delay, especially with an asymmetric traffic volume.

Cooperative collision avoidance with imperfect vehicle-to-infrastructure (and vice-versa) communication is analyzed in \cite{Colombo2015}. The centralized supervisor, located at the intersection, acquires the positions, velocities, and accelerations of the incoming vehicles, and then decides either to allow vehicles' desired inputs, or to override them with a safe set of inputs. The communication is subject to failures, with the success reception probability based on the Rayleigh fading model. According to their simulation results, the mean time between the accidents is significantly increased, but a collision may happen if the override message has been lost.

A hybrid centralized/distributed architecture that ensures both safety (no collisions), and liveness (a finite crossing time), at intersections without stop signs and traffic lights, is proposed in \cite{Kowshik2011}. The vehicles are equipped with a positioning unit, internal sensors, and a V2V communication unit. To resolve the problem with a bounded communication delay and packet losses, the rear car needs to break with maximum deceleration. They compared the proposed solution with stop-sign and traffic-light technologies and found that the average travel time is significantly reduced.

\subsection{Our contributions}\label{subsec:ourcontr}
Although state-of-the-art provide solutions for many different problems, to the best of our knowledge, there is no solution that can handle an unknown number of communication failures. For instance, the solution in \cite{Colombo2015} cannot handle the failure in the override message, while solutions in \cite{Hafner2013, Kowshik2011} can handle only a predefined communication delay. The solutions in \cite{Azimi2013} count on other sensors to avoid collision in the presence of communication failures. In contrast to these solutions, we aim to provide a solution that can handle an unknown and large (yet finite) number of communication failures, and preserve other important characteristics of the state-of-the-art methods such as distributed implementation and position-aware decisions. We also provide an analysis that show that both safety and liveness requirements are satisfied, and the numerical results, based on a real data set, that show that the crossing delay is only slightly increased even in the presence of highly correlated failures.

\section{Problem formulation}\label{sec:smodel}
\begin{figure}[!t]
\centerline{
\includegraphics[width=0.7\columnwidth]{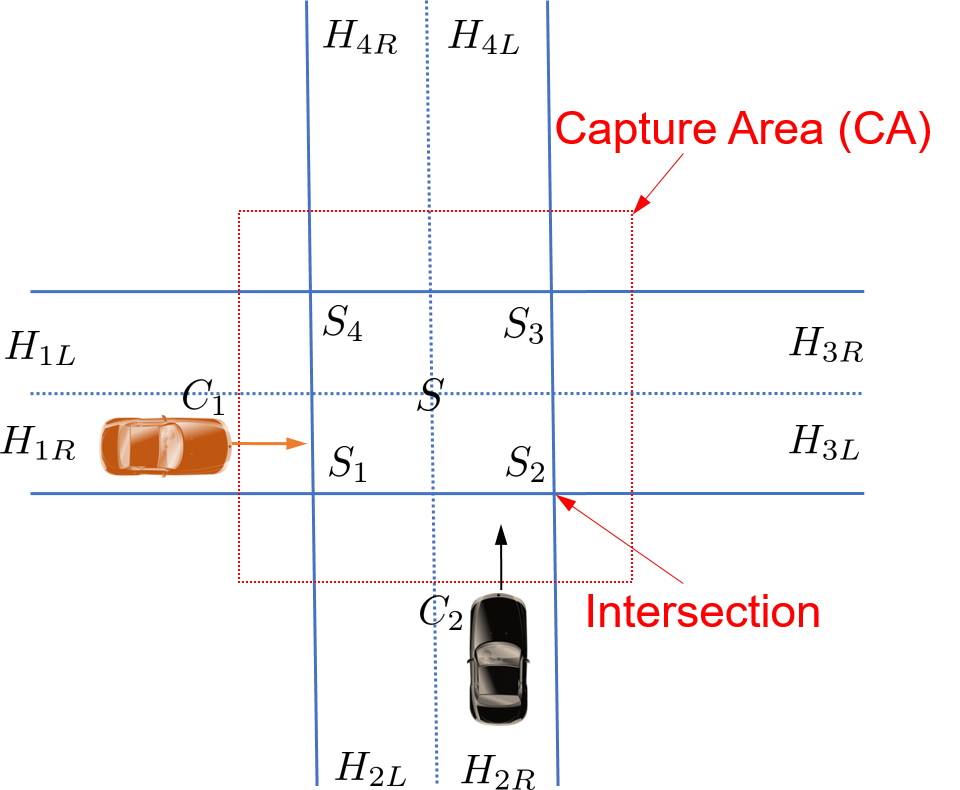}}
\caption{An illustration of 2-lane road intersection with two incoming cars.}
\label{fig:intersection2cars}
\end{figure}

We consider two fully autonomous cars ($C_1$, $C_2$) on different lanes, competing to cross a road intersection as depicted in Fig. \ref{fig:intersection2cars}. They are equipped with V2V communication unit, GPS or DGPS for position estimation, and an IMU for velocity/acceleration estimation \cite{Skog2009,Vu2013}. Moreover, they have an equal size and weight, and unique identifier. Initially, no car has a priority to cross the intersection. Both cars are moving towards the intersection, and when necessary they can slow-down/speed-up with a constant acceleration/deceleration. We neglect the velocity and acceleration errors, but not the position errors since they may be large.\footnote{These assumptions are made to facilitate the presentation of the main idea, but an extension to more complex models is possible.}

The road has 2 lanes, so the intersection ($S$) can be divided into four subsections ($S_1, S_2, S_3, S_4$). In each of these subsections, collision may occur if two cars occupy it simultaneously. Since cars' acceleration is limited by their inertia, we also define a \textit{capture area (CA)}, i.e., the area in which no control action can stop the entrance to the intersection. Note that the capture area is not constant, and it depends on the cars' dynamics.

\subsection{Position-related information}\label{subsec:is-model}
Let us define the following variables:
\bi
\item[] $UID_j$ - unique identifier of $C_j$,
\item[] $x_{j}^t$ - true longitudinal (1D) position of $C_j$ at time $t$,
\item[] $\hat{x}_{j}^t$ - estimated longitudinal (1D) position of $C_j$ at time $t$,
\item[] $\sigma_{x,j}$ - standard deviation of the position estimate of $C_j$,
\item[] $v^t_{j}$ - true longitudinal (1D) velocity of $C_j$ at time $t$,
\item[] $a^t_{j}$ - true longitudinal (1D) acceleration of $C_j$ at time $t$,
\item[] $\Theta^t_j$ - set with absolute (2D) position, velocity, and acceleration of $C_j$ at time $t$,
\item[] $x_{S}$ - the central point of the intersection $S$,
\item[] $CLANE_j$ - current lane, before crossing the intersection ($CLANE_j
\in \{H_{1R}, H_{2R}, H_{3R}, H_{4R}\}$),
\item[] $NLANE_j$ - next lane, after crossing the intersection ($NLANE_j
\in \{H_{1L}, H_{2L}, H_{3L}, H_{4L}\} \backslash H_{jL}$).
\ei
Time index $t=1,\ldots,N_t$ represents the discrete time slot, and the time interval between two time slots is denoted with $T$. Both positioning and the distributed IC algorithm uses the same time slot. Note that the time indexes are omitted for variables that remain constant with time.

We assume that cars periodically (with period $T$) broadcast a \textit{heart-beat (HB)} message. This message (to be defined in the next section) is transmitted in all empty time slots (i.e., when there are no other messages), to ensure that both cars can detect each other. Once car $C_j$ ($j=1,2$) gets close enough to the capture area, it sends the 'ENTER' message. Since this is a safety-critical problem, this message will be sent as soon as the following condition is satisfied:
\be\label{eq:sendEnter}
{\rm{COND1}}: ~~~P_{j,CA}^{t}=Prob~\{x^{t+t_{j,1}}_j\in CA\} \ge \epsilon
\ee
where $\epsilon$ is the desired tolerance (e.g., $\epsilon=10^{-9}$), and $t_{j,1}$ is the number of time slots before $C_j$ gets the intersection. This number should be set to the value that would allow car to start communication as soon as it is within the communication range ($R$) of another car. For example, given the current velocity ($v^t_{j}$), and assuming zero acceleration ($a^t_{j}=0$), we can set $t_{j,1}=\lceil R/(v^t_{j}\cdot T)\rceil$ where $\lceil~\rceil$ is the ceiling operator.

Once car $C_j$ crosses the intersection, it sends the 'EXIT' message, and this will happen once:
\be\label{eq:sendExit}
~~~~~{\rm{COND2}}: ~~~P_{j,N}^{t}=Prob~\{x^{t}_j\in NLANE_j\} \ge 1-\epsilon 
\ee
The probability $P_{j,CA}^{t}$ needs to be computed at each time slot before $C_j$ sends the 'ENTER' message, and the probability $P_{j,N}^{t}$ only after $C_j$ decides to cross the intersection. These probabilities can be computed from the predictive probability distribution, and posterior probability distribution, which can be found via Kalman or Particle filtering \cite{Arulampalam2002}. %

Now we define the parameter that will be used to determine the priority for intersection crossing. One may assign the priorities a priori (e.g., via \textit{UIDs}, such as in \cite{Azimi2013}) based on the type and the importance of the car (e.g., a police car would go first), but this would cause an additional delay. We instead use the current position estimate and the cars' dynamics to compute the \textit{mean time to intersection (MTI)}. The MTI of $C_j$ at time $t$ is given by:
\be\label{eq:mti}
\tau^t_{{\rm{MTI}},j}=\frac{-v^t_{j}+\sqrt{(v^t_{j})^2+2a^t_{j}({x}_{S}-\hat{x}^{t}_j)}}{a^t_{j}}
\ee
The car with lower MTI will first cross the intersection, while the other car would need to wait for the 'EXIT' message. In the rare situation, in which MTIs are equal, we use instead \textit{UIDs} as a tie-breaker. Note that the priority management is not a safe-critical operation, so we do not need to consider the variance of the GPS estimate. 

\subsection{Message format and failures}\label{subsec:msgs}
The messages should include all relevant information required for safe and efficient IC. We use here a similar set of messages as in \cite{Azimi2013}, which are defined according to DSRC SAE J2735 standard \cite{sae2009}. To adapt to our problem, we make three modifications: (i) we do not transmit the data not needed for IC (such as trajectory list), (ii) we do not transmit a 'CROSS' message since the crossing time interval is implicitly available from other messages, and (iii) we introduce a 'HB' message in order to handle failures.

The format of the messages is given as follows:
\bn
\item 'HB' message: $~MSGHB^t_j=\{UID_j,MSGTYPE_j,\Theta^t_j\}$
\item 'ENTER' message: $MSGENTER^t_j=\{UID_j,MSGTYPE_j,CLANE_j,NLANE_j,\tau^t_{{\rm{MTI}},j}\}$
\item 'EXIT' message: $~MSGEXIT^t_j=\{UID_j,MSGTYPE_j,NLANE_j\}$

\en
where $MSGTYPE_j \in \{{\rm{'HB'}},{\rm{'ENTER'}},{\rm{'EXIT'}}\}$. We also make the following assumptions: 
\bi
\item Cars can experience an \textit{unknown} number of consecutive receive-omission failures (i.e., fail to receive the message). Without loss of generality, we consider one burst of errors, and denote it by $f_j$ ($f_j \ge 0$) for car $C_j$.
\item Cars will eventually (i.e., in round $f_j+1$) \textit{succeed} to receive the sent 'ENTER' and the 'EXIT' messages.
\item The 'HB' message must be received \textit{at least once} before the IC algorithm starts.
\item Each message ('HB', 'ENTER', or 'EXIT') is sent \textit{within one packet}, so any of them will be either fully delivered, or completely lost.
\item If the transmitted message is not received in the same time slot, it is considered outdated and discarded.
\item Cars are able to successfully transmit all messages, and the delivered packet does not contain erroneous data.
\item Cars are fully cooperative and they never send malicious messages. 
\ei

Based on these assumptions, we focus on the most frequent failures caused by obstructed wireless channel (e.g., non-line-of-sight, jammers, interference). We also do not make any assumption about the channel model (such as Rayleigh fading \cite{Colombo2015}), nor predefine the number of failures. However, we do not consider send-omissions, nor erroneous data, since these problems are highly unlikely with a well-tested equipment and an appropriate error-correcting code \cite{Kowshik2011}. Regarding 'HB' messages, since they are sent in each time slot before the 'ENTER' message, and that communication range is typically large (few hundred meters), it is reasonable to assume that at least one of them will be received.\footnote{If this condition is not satisfied, the communication link is permanently damaged and the cars need to rely on other sensors (see Section \ref{subsec:relwork}), such as radar or camera. Therefore, it is strongly advisable to have multiple technologies that operate independently.} Other problems, such as malicious behavior, are out of focus of this paper, but can be partially resolved using another type of algorithms \cite{Raynal2010}.

\subsection{Control actions}\label{subsec:msgs}
We define here control actions that should be performed in the presence of failure ($SAFECTRL$), and after the agreement is established ($MAINCTRL$). 

Once the distributed algorithm is performed, both cars have access to each others 'ENTER' messages. Therefore, using $\tau^t_{{\rm{MTI}},j}$, $CLANE_j$ and $NLANE_j$, both cars can determine if collision is possible. There are two situations in which the collision cannot happen: (i) cars never occupy the same subsection, and (ii) cars do not occupy the same subsection simultaneously. Otherwise, the collision is likely to happen. The collision area ($COL$) depends on the cars' routes and may be any of the subsections ($S_1, S_2, S_3, S_4$) or a combination of them. It is also possible that the collision area is empty ($COL = \emptyset$), e.g., if both cars intend to turn right.

Therefore, $MAINCTRL$ should let $C_1$ to proceed with the desired acceleration if $COL = \emptyset$ or $\left|\tau^t_{{\rm{MTI}},1}-\tau^t_{{\rm{MTI}},2}\right|>\tau_{TH}$ where $\tau_{TH}$ is the threshold that depends on cars' velocity, and the minimum safety distance. Otherwise, the collision is possible, so $C_1$ can proceed either if $\tau^t_{{\rm{MTI}},1}<\tau^t_{{\rm{MTI}},2}$ or $\tau^t_{{\rm{MTI}},1}=\tau^t_{{\rm{MTI}},2}$ \& $UID_1>UID_2$. As we can see, the identifiers are used as a tie-breaker in the rare circumstances in which the MTIs are exactly the same.

Whether $C_j$ ($j=1,2$) has a priority or a collision is not possible, it can keep moving with the desired acceleration $a_{j,PR}$, for example, equal to the current acceleration:
\be\label{eq:pr}
a_{j,PR}=a_j^{t+1}=a_j^{t+2}=\ldots = a_j^{t}
\ee
Otherwise, it needs to slow down just little bit to avoid collision. Assuming that we want to reduce cars' displacement for $D$ (which should be at least equal to the width of the $COL$), this acceleration ($a_{j,NOPR}$) can be found using standard kinematic equations:
\be\label{eq:nopr}
a_{j,NOPR}=a_j^{t+1}=a_j^{t+2} = \ldots = a_j^{t}-\frac{2D}{(\tau^{t+1}_{j,COL})^2}
\ee
$\tau^{t+1}_{j,COL}$ is the worst-case remaining time to the collision area (assuming constant acceleration), and is given by:
\be\label{eq:timeCol}
\tau^{t+1}_{j,COL}=\frac{-v^{t+1}_{j}+\sqrt{(v^{t+1}_{j})^2+2a^{t+1}_{j}({x}_{COL}-\hat{x}^{t+1}_{j,MAX})}}{a^{t+1}_{j}}
\ee
where ${x}_{COL}$ is the entrance point of the collision area, and $\hat{x}^{t+1}_{j,MAX}$ is near-worst-case position estimate of $C_j$ (e.g., $\hat{x}^{t+1}_{j,MAX}=\hat{x}_j^{t+1}+l\cdot \sigma_{x,j}$ with $l \ge 3$). Note that this computation needs to be done at time $t$, so predictive probability distributions are required. Once $C_j$ receives the 'EXIT' message, it will increase the acceleration to the desired value $a_{j,EXIT}>a_{j,NOPR}$, e.g., to the same value as in \eqref{eq:pr}.

Now we define $SAFECTRL$, which should be performed in the presence of communication failure. Once $C_j$ becomes aware of its own or other's car failure, it needs to decelerate fast enough so that it can stop before entering the $COL$. This can be ensured by setting the acceleration to the value that would ensure zero velocity at the entrance of the $COL$:
\be\label{eq:safe}
a_{j,SAFE}=a_j^{t+1}=a_j^{t+2}= \ldots = -\frac{v^{t+1}_{j}}{\tau^{t+1}_{j,COL}}
 \ee
Note that $a_{j,SAFE}$ is feasible since the $C_j$ is by assumption out of the CA. Once the failure problem is resolved, $MAINCTRL$ can be again executed. Consequently, the car that stayed longer in this state (with sharp deceleration) will lose the priority. In case of too many failures, both cars would stop, so $(\tau^t_{{\rm{MTI}},1},\tau^t_{{\rm{MTI}},2}) \rightarrow (\infty,\infty)$, and the $UID$s would be used to choose the priority.

The integration of these control actions within the distributed IC algorithm are provided in the following section.

\section{Distributed IC algorithm in the presence of communication failures}\label{sec:distalg}
Given the models from the previous section, we now propose a distributed IC algorithm that ensures safe and efficient intersection crossing in the presence of unknown number of communication failures.

\subsection{Algorithm description}\label{subsec:algDescr}

\begin{figure*}[!th]
\centerline{
\includegraphics[width=0.99\textwidth]{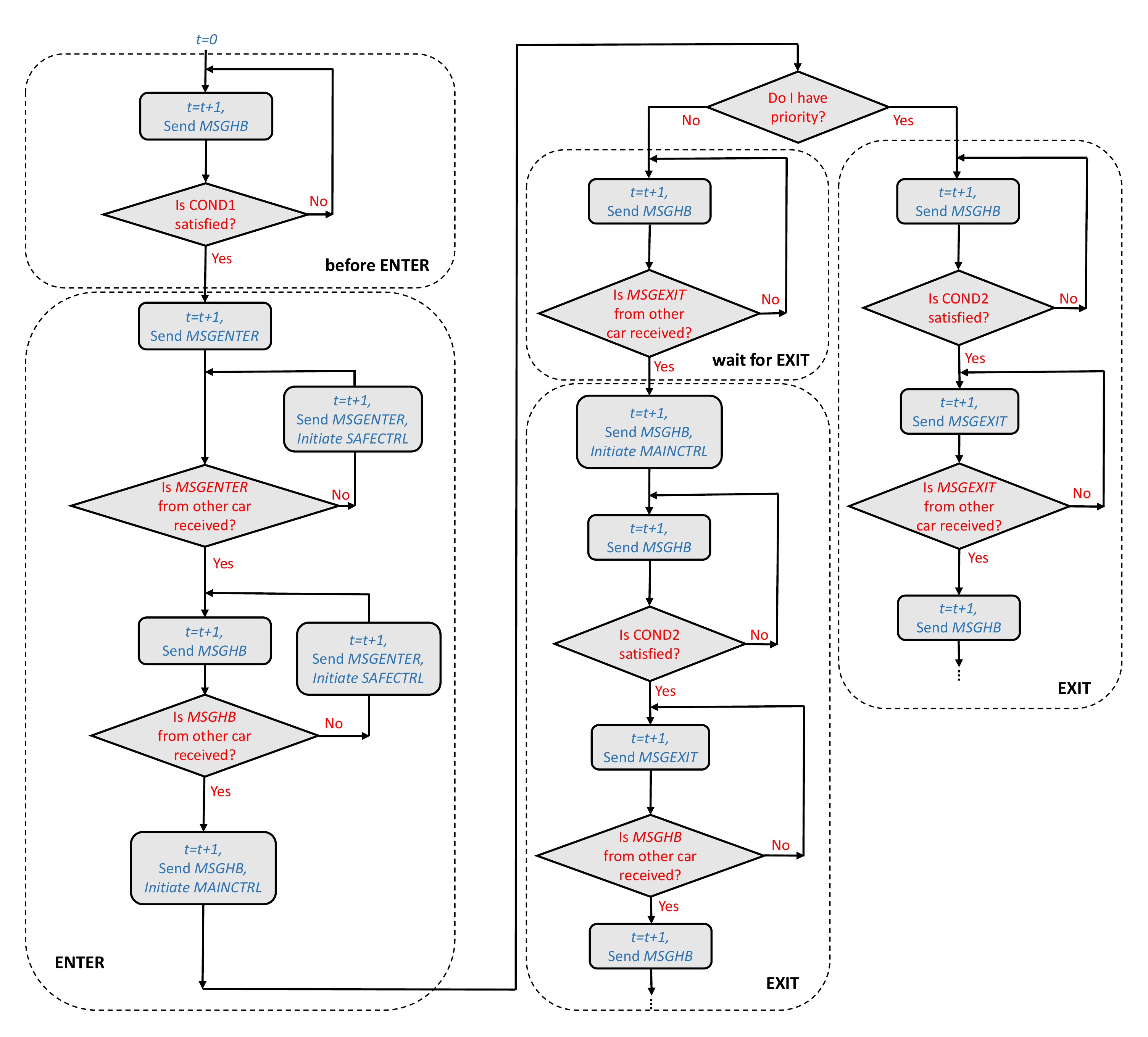}}
\caption{Flowchart of the IC algorithm for car $C_j$. The time and car indexes are omitted for ease of presentation.}
\label{fig:flowchartIC}
\end{figure*}

The flow-chart of the algorithm, shown in Fig. \ref{fig:flowchartIC}, consists of the following modules: (i) before ENTER, (ii) ENTER, (iii) wait for EXIT (only for the car without priority), and iv) EXIT. Here we provide the description of these modules:
\bn
\item \textbf{Before ENTER}: This module is a simple event detector, in which car waits for COND1 to be satisfied. Once that happen, the car is aware that it will soon reach the intersection and may collide with another car.
\item \textbf{ENTER}: Once COND1 is satisfied for $C_j$, this car will attempt to send the 'ENTER' message, and then check if the same message from the other car is received. This messages may not be received either because of the failure, or because other car is still waiting for COND1 to be satisfied. In that case, $C_j$ will repeat sending of the 'ENTER' message and initiate $SAFECTRL$. Then, once the 'ENTER' message from the other car is received, $C_j$ will check if the 'HB' message is received. This message serves as an acknowledgment that the 'ENTER' message is received by the other car. If it is not received, $C_j$ will repeat it, and initiate $SAFECTRL$. Otherwise, $C_j$ has already received the 'ENTER' message, and is aware that other car has received its 'ENTER' message. Therefore, both cars will initiate $MAINCTRL$ action.
\item \textbf{Wait for EXIT}: This module is only executed for the car without priority. This car will wait for the other car to execute $MAINCTRL$, exit the intersection, and confirm it by sending the 'EXIT' message. 
\item \textbf{EXIT}: While executing $MAINCTRL$, the car with priority will wait for COND2 to be satisfied, then send the 'EXIT' message to the other car, and check if the 'EXIT' from the other car is received. The 'EXIT' message will permit the car without priority to execute $MAINCTRL$, check if COND2 is satisfied, then send its 'EXIT' message, and finally, check if the other car has received it. The 'EXIT' messages are repeated in case of failures, but there is no need to execute $SAFECTRL$ since collision is now impossible. 
\en

Note that the cars also execute other tasks simultaneously (e.g., an algorithm for pedestrian detection or rear-end collision avoidance) that are not shown in this algorithm, and they also may trigger $SAFECTRL$. Note also that control actions ($MAINCTRL$ and $SAFECTRL$) are only initiated within one time slot, but their full execution will take much longer.

\subsection{Time diagrams}\label{subsec:timeDiag}

\begin{figure*}[!t]
\centerline{
\includegraphics[width=0.99\textwidth]{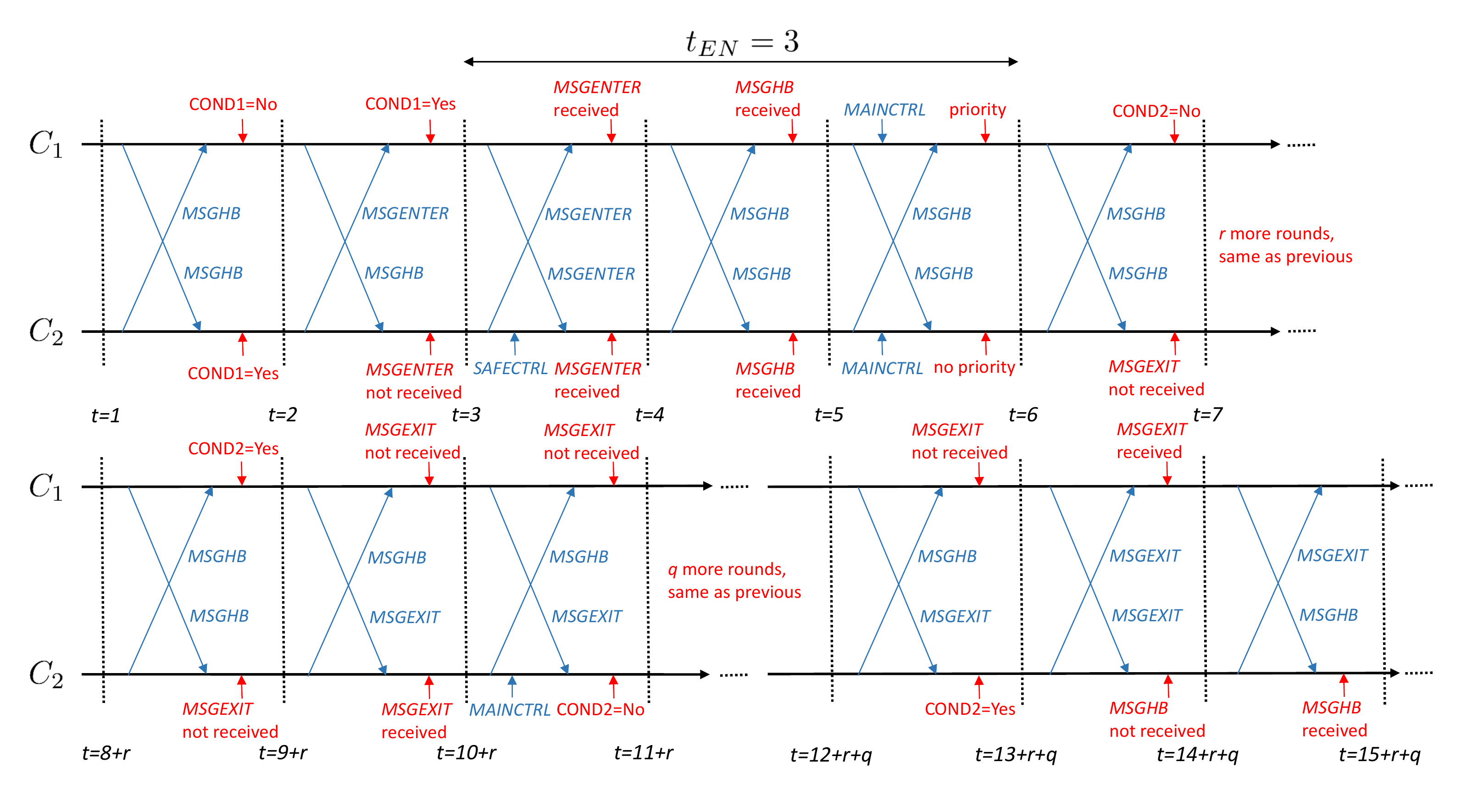}}
\caption{The time diagram of the full IC algorithm without failures.}
\label{fig:diagramFull}
\end{figure*}

We start with the example of time diagram (Fig. \ref{fig:diagramFull}) for the execution without failures. We note that crossing the intersection will take many rounds, which depends on the cars' dynamics. For instance, if communication round takes 100 ms, and the crossing time takes 3 s, the crossing would take 30 communication rounds. Therefore, any failure would just slightly increase the total delay. We can also see the $MAINCTRL$ is executed in the same round ($t=5$), when both cars have available both 'ENTER' messages. $SAFECTRL$ is initiated by $C_2$ because $C_1$ sent its 'ENTER' message one round later. However, since this action is overwritten by $MAINCTRL$ just two rounds later, there would not be enough time for a noticeable deceleration. After initiating $MAINCTRL$, $C_1$ would get priority (as an example), while $C_2$ would need to slow down little bit to avoid collision. Then, after $C_1$ exits the intersection, it will send the 'EXIT' message, and keep repeating it until it gets the 'EXIT' message from $C_2$ (which will happen once $C_2$ exits the intersection). Finally, in the last two rounds, both cars become aware that the other car performed the required actions.

\begin{figure*}[!th]
\centerline{
\subfloat[]{\includegraphics[width=0.77\textwidth]{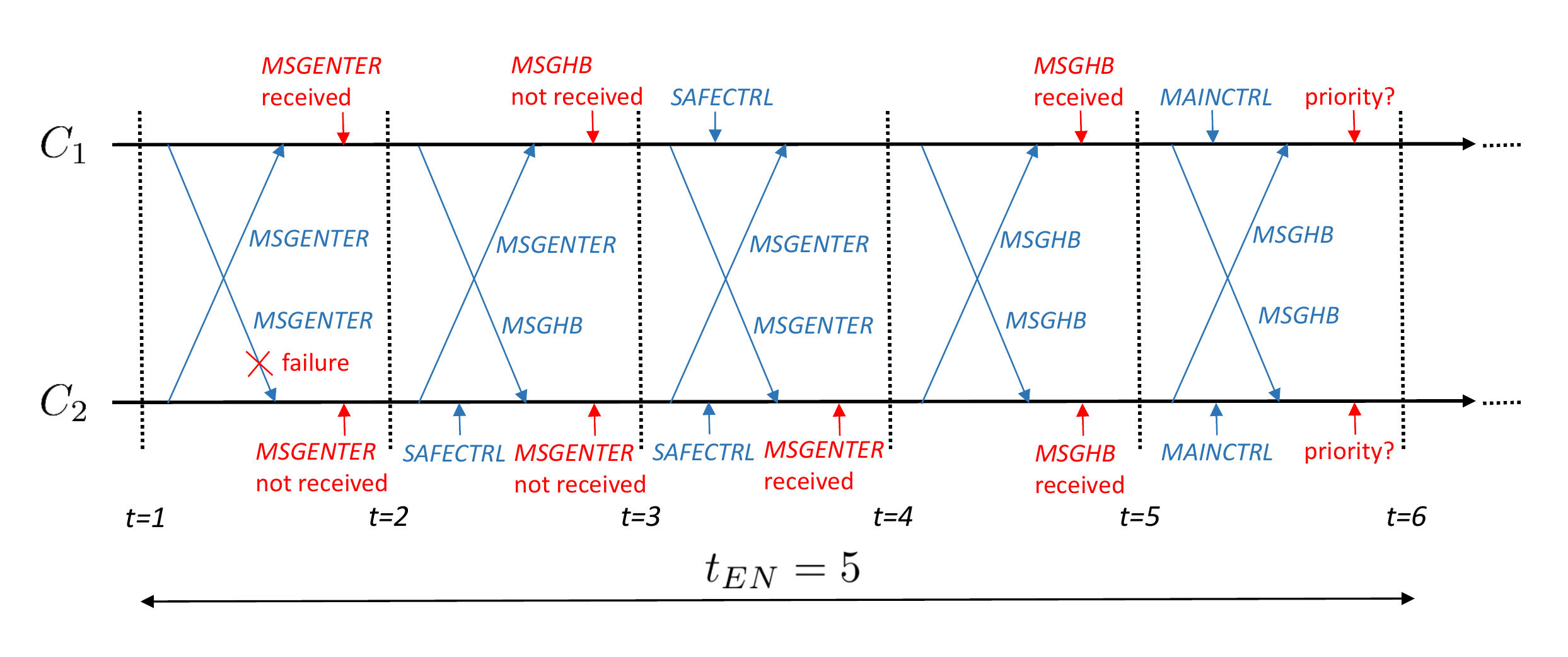}\label{fig:diagram01fail}}
}
\centerline{
\subfloat[]{\includegraphics[width=0.99\textwidth]{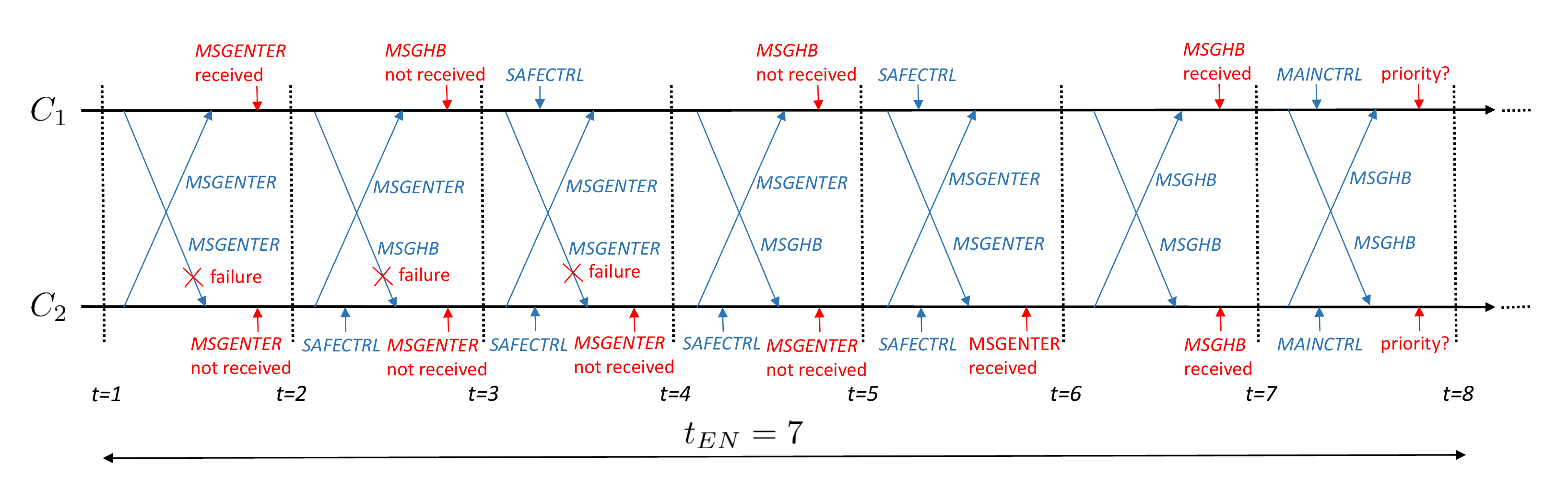}\label{fig:diagram03fail}}
}
\caption{Time diagrams of the 'ENTER' part of the IC algorithm for different number of failures: (a) $(f_1, f_2)=(0,1)$, (b) $(f_1, f_2)=(0,3)$}. 
\label{fig:diagramsFail}
\end{figure*}

We now analyse the examples with communication failures. As shown in Fig. \ref{fig:diagramsFail}, we consider the following situations: $(f_1, f_2)=(0,1)$ and $(f_1, f_2)=(0,3)$. We focus only on the ENTER part of the algorithm, since this is the most critical part in which a collision may happen. According to Fig. \ref{fig:diagramFull}, this part takes $t_{EN}=3$ rounds without failures.

In the first example (Fig. \ref{fig:diagram01fail}), $C_2$ fails to receive once the 'ENTER' message sent by $C_1$. Therefore, it will initiate $SAFECTRL$ and repeat the 'ENTER' message. Meanwhile, $C_1$ received the 'ENTER' message, so it can transmit the 'HB' message to confirm it. $C_1$ also expects to receive 'HB' message from $C_2$, but it will receive 'ENTER' instead, and figure out that there is a failure at $C_2$. Consequently, $C_1$ will send again 'ENTER' message and initiate $SAFECTRL$. This message will be received by $C_2$, so it can transmit the 'HB' message. The same message is also sent by $C_1$ in the same round. Then, both cars will execute $MAINCTRL$ in the same round and decide about the priority. The total delay in this example is $t_{EN}=5$. 

In the second example, $C_2$ fails to receive the message for three consecutive rounds. We note that in the second round 'HB' message is not received by $C_2$ in contrast to previous example. Since this message is not needed, the second failure would not cause extra rounds. However, the third failure will cause extra two rounds since a new 'ENTER' message will not be available in the round following this failure. Then, the last four rounds of the diagram are the same as in the previous example, and the total delay is $t_{EN}=7$. 

In summary, the total delay depends only on the maximum number of failures, and only an odd failure (3, 5, 7, etc.) increases the delay for two extra rounds. Therefore, it follows (by induction) that the total delay of the ENTER part is given by: $t_{EN}=2 \lceil\max{(f_1,f_2)}/2\rceil+3$.

\section{Numerical results}\label{sec:numres}

Our goal is to analyze the delay caused by communication failures. For that purpose, we use the real measurements of the packet delivery ratio ($P_{PDR}$), available in \cite{Bai2006}. In this work, authors analyzed 802.11p based DSRC communication for V2V communication, and characterized communication and application level reliability of the wireless channel. They used General Motors cars equipped with a DSRC radio, omni-directional antenna and a GPS receiver. The transmission power was 20 dBm, the communication range was about 500 m, and the sampling interval was 100 ms. The experiments were conducted on GM test freeways under open-field environment (without any obstacles) and a realistic harsh environment (with many obstacles such as tunnels and bridges). 

\begin{figure*}[!tb]
\centerline{
\subfloat[]{\includegraphics[width=0.49\textwidth]{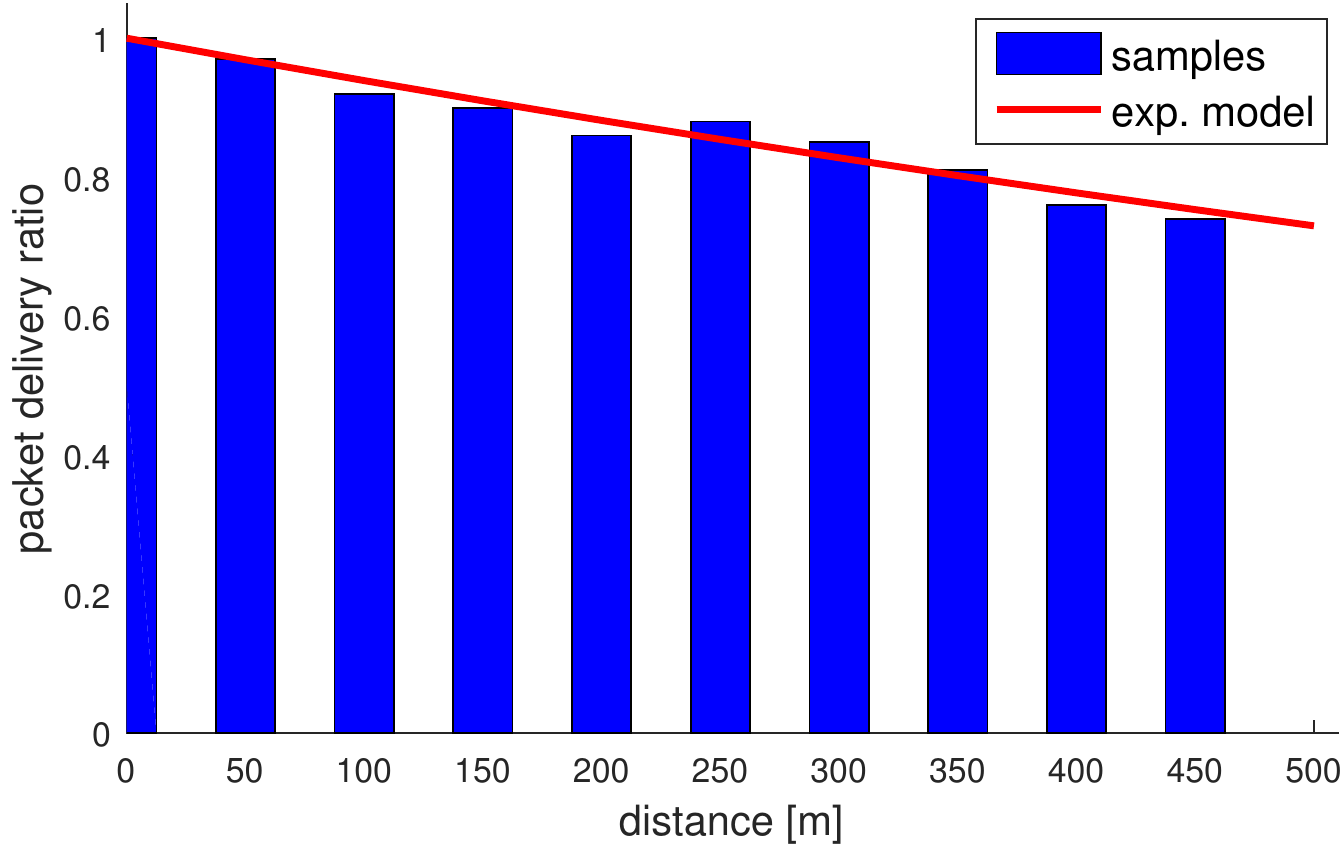}\label{fig:failDistOpen}}
\subfloat[]{\includegraphics[width=0.49\textwidth]{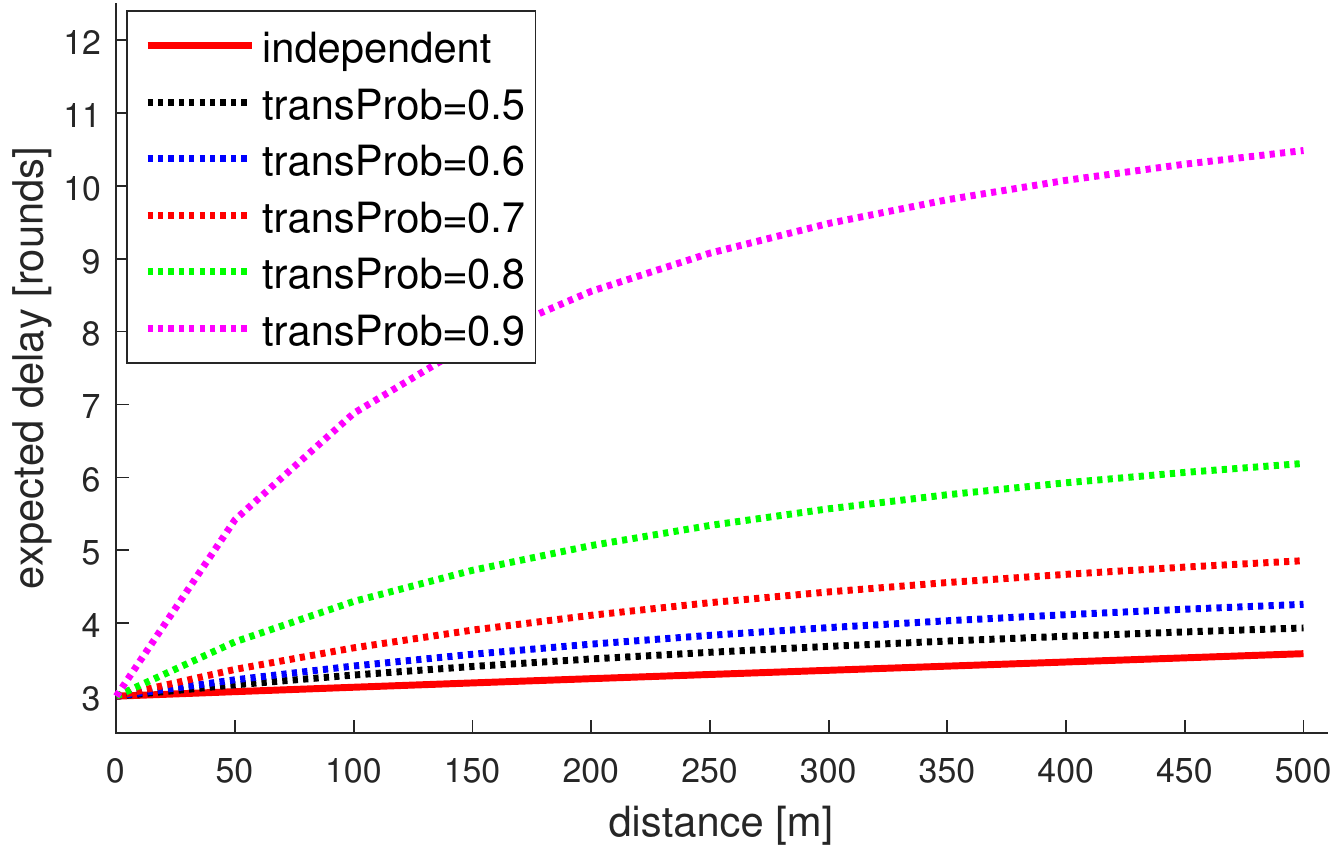}\label{fig:delayDistOpen}}
}
\centerline{
\subfloat[]{\includegraphics[width=0.49\textwidth]{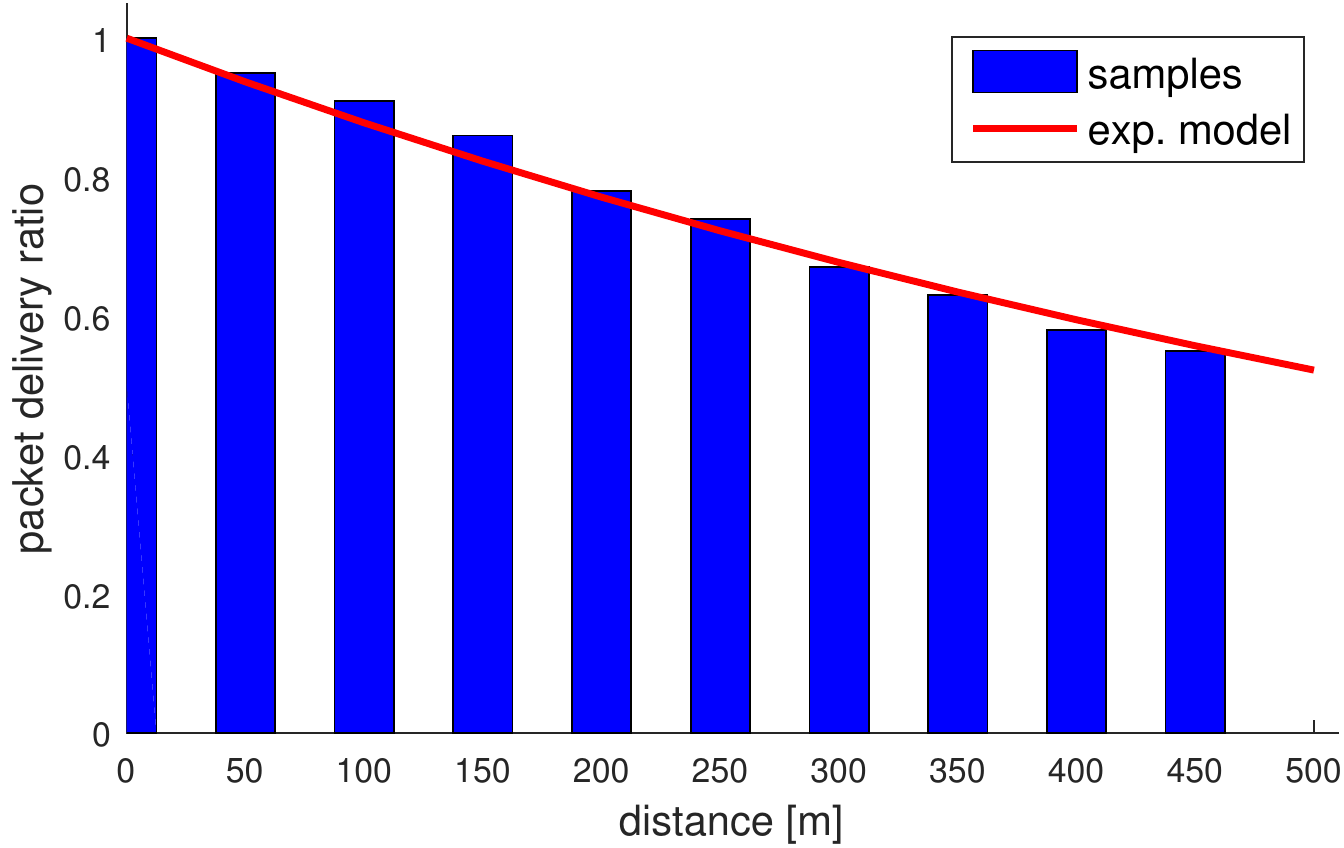}\label{fig:failDistFree}}
\subfloat[]{\includegraphics[width=0.49\textwidth]{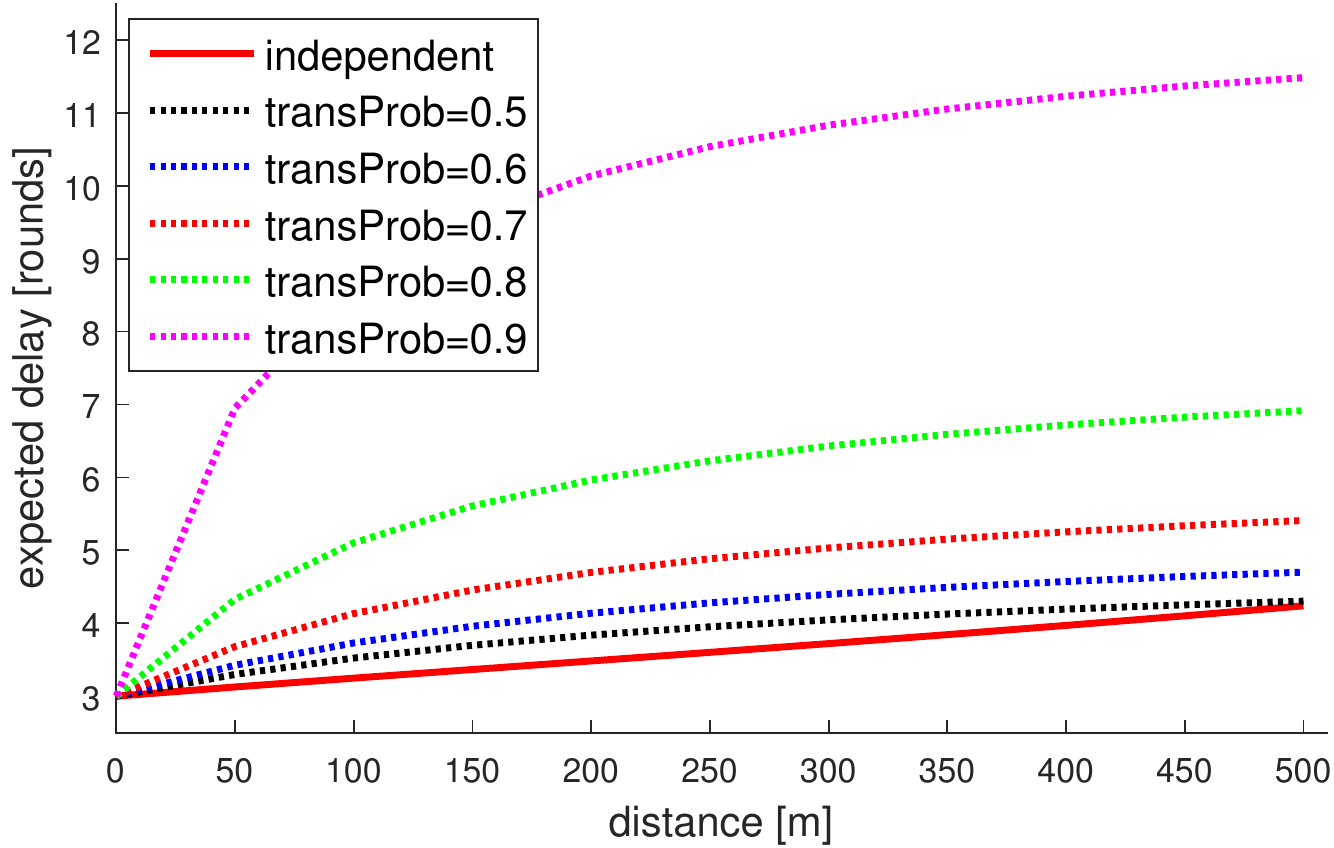}\label{fig:delayDistFree}}
}
\caption{(a) Packet delivery ratio vs distance (open-field), (b) Expected delay vs distance (open-field), (c) Packet delivery ratio vs distance (harsh), (d) Expected delay vs distance (harsh). }
\label{fig:FailuresDelays}
\end{figure*}

We use the measurements of $P_{PDR}$ as a function of distance between the cars, for both open-field and harsh environment. Then, we used exponential model ($e^{-\lambda d}$, where $\lambda$ is the decay rate [$m^{-1}$], and $d$ is the distance [$m$]) to model $P_{PDR}$ as a function of distance. The results are shown in Fig. \ref{fig:failDistOpen} and Fig. \ref{fig:failDistFree}, and the corresponding decay rates are 0.00063 and 0.0013. 

We then compute the expected delay of the ENTER part of the algorithm ($\hat{t}_{EN}$) by averaging over different number of communication failures. We consider the scenario in which one car commits $f$ consecutive failures, while other car commits no failures. Assuming independence, the likelihood of $m$ consecutive failures is given by geometrical distribution: $p(f=m) = (1-P_{PDR})^m \cdot P_{PDR}$, so we can compute the delay as follows:
\be\label{eq:expectedDelay}
\hat{t}_{EN}=\frac{\sum_{m=0}^{M}{p(f=m) \cdot t_{EN}(f=m)}}{\sum_{m=0}^{M}{p(f=m)}}
\ee
where $M$ is the maximum number of failures (we set it to $M=50$ since more failures would cause a negligible delay). However, although the independence assumption is experimentally justified in \cite{Bai2006}, it may not be the case if there is a long obstruction of the channel (e.g., due to the large truck in front of the car). For this case, we define a transitional probability $\xi=p(f=m|f=m-1)$ which gives us an information about the likelihood of the failure in the current round, given that failures already happened in the previous round. The likelihood of $m$ consecutive failures is now given by: $p(f=m) = (1-P_{PDR}) \cdot P_{PDR} \cdot \xi^{m-1}$, and $\hat{t}_{EN}$ can be again computed using \eqref{eq:expectedDelay}.

The results of $\hat{t}_{EN}$ as a function of distance for different values of are shown in Fig. \ref{fig:delayDistOpen} and Fig. \ref{fig:delayDistFree} for both open-field and harsh environment, respectively. As expected, the open-field environment leads to consistently lower delay, but the difference is not significant ($\le 20\%$). However, high transitional probabilities can cause a significant delay, especially for $\xi=0.9$, but this delay is still much lower comparing with the total crossing time that typically takes few seconds. However, in highly unlikely scenario in which $\xi$ is too close to 1, the delay would be too large (infinite, for $\xi=1$), so in that case an alternative technology should be used.

\section{Conclusions}\label{sec:conc}
We proposed a novel IC algorithm that can handle an unknown nymber of communication failures. In order to avoid collision and minimize the delay, our algorithm uses the cars' positions and their dynamics to adapt their actions. The algorithms is fully distributed, so no centralized intersection manager is required. We provided an analysis of time diagrams that show that both safety and liveness are satisfied in all realistic situations. According to our numerical results, which are based on real measurements, the crossing delay is just slightly increased even in the presence of correlated failures. Our future work will focus on the extension of this algorithm for a variable number of cars, and its implementation within standardized traffic simulators. One may also consider other challenging problems, such as development of a rear-end collision avoidance algorithms in the presence of communication failures, and development of an IC algorithm that can handle malicious massages.

\section*{Acknowledgment}
This work was supported by Area of Advance Transport, Chalmers University of Technology.

\footnotesize
\bibliographystyle{ieeetr} 
\bibliography{paper-ic-refs}

\end{document}